\documentclass{article}

\usepackage{arxiv}

\usepackage[utf8]{inputenc} 
\usepackage[T1]{fontenc}    
\usepackage{hyperref}       
\usepackage{url}            
\usepackage{booktabs}       
\usepackage{amsfonts}       
\usepackage{nicefrac}       
\usepackage{microtype}      
\usepackage{graphicx}
\usepackage{doi}
\usepackage{setspace}
\usepackage{changes}
\usepackage{amsmath}
\usepackage{float}
\usepackage{xcolor}
\title{Semianalytical modeling of the mass transfer in microfluidic electrochemical chips}

\date{\textit{Revised on \today}}

\author{ \href{https://orcid.org/0000-0002-4614-1867}{\includegraphics[scale=0.06]{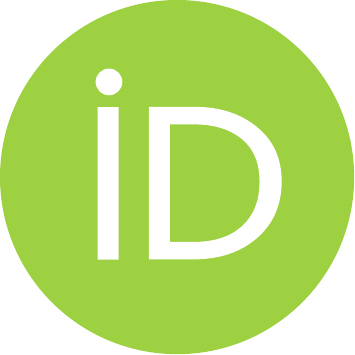}\hspace{1mm}Stéphane Chevalier}\thanks{Corresponding Author: Prof. Stéphane Chevalier, Esplanade des Arts et Métiers, 33405 Talence Cédex, FRANCE} \\
	Arts et Métiers Institute of Technology\\
	Esplanade des Arts et Métiers\\
	33405 Talence \\
	\texttt{stephane.chevalier@u-bordeaux.fr} \\
}

\renewcommand{\shorttitle}{\textit{Published in PRE}}

\hypersetup{
pdftitle={Semianalytical modeling of the mass transfer in microfluidic electrochemical chips},
pdfsubject={orignal research paper},
pdfauthor={S. Chevalier},
pdfkeywords={Microfluidic channel, Electrochemical reaction, Mass diffusion ,  Current density distribution , Analytical model , Fuel cells},
}

\begin{document}
\maketitle

\begin{abstract}
This paper reports a mass transfer model of a reactant flowing in a large aspect ratio microfluidic chip made of a channel with electrodes on the side walls. A semi-analytical solution to the two-dimensional Fickian diffusion of a reactant in a microchannel, including the electrochemical reaction at the electrode interface, and the velocity profile obtained from the Navier-Stokes equations in a fully developed laminar regime is found . The solution is written in the Laplace domain in terms of transfer functions. The proposed solution is an extension of the Lévêque approximation describing the reactant diffusion from the electrode to the middle of the microfluidic channel. The main applications of this work are the use of the obtained transfer functions for the measurement of the Faradic current density or the chemical concentration at the electrode interface. The study can also be extended to the heat transfer in microfluidic electrochemical chips (temperature or heat flux measurements at the electrode interface).
\end{abstract}

\vspace{2cm}

\clearpage

%
%
%
%

\section{Introduction}
A microfluidic electrochemical chip is made of a microchannel (width and height typically lower than a millimeter), with electrodes on opposite walls (either top and bottom or left and right). An electric potential and a current density is generally applied between these electrodes. In such devices, several chemicals flow to produce a charge, mass, and often heat transfer at the electrode interface. Due to their small dimensions and the good control of the hydraulic conditions, they are more frequently used for a wide range of applications. For example, they are good candidates to improve chemical energy conversion into electricity \cite{Modestino2016,Esposito2017}. In their comprehensive review, Safdar et al. \cite{Safdar2016} reported that microfluidic fuel cells have unique advantages compared to the conventional fuel cells such as high surface area-to-volume ratio, ease of integration, cost effectiveness, and portability . Another major application of the microfluidic electrochemical device is the electrodialysis \cite{Nguyen2020}, in particular to produce drinking water from sea water. In their works, Schlumpberger et al. \cite{Schlumpberger2015} reported that extensive research is underway to develop improved water treatment methods, which has attracted renewed attention in such applications. This confirms the unique capabilities of microfluidic electrochemical chips towards the development of new technologies. Microfluidic elecrochemical chips also embed a wide range of sensors for biological and medical diagnostic such as glucose monitoring  \cite{Didier2020}. Thus, following these remarkable efforts to develop new and efficient microfluidic electrochemical chips, the need for a better understanding of the mass transfer between the electrodes in microfluidic channels is crucial to predict the performances of these devices and to specify new guidelines for improved chip designs. In particular, the combination of the velocity profile and the mass diffusion has to be taken into account properly to describe the correct electrode mass transfer, or eventually the heat transfer as it is governed by similar physics and occurs simultaneously during any chemical reaction \cite{Ryu2017}. \\

The heat and mass diffusion in microchannels have been extensively studied through the literature \cite{Galfi1988,Baroud2003,Uema2021}. Most of the studies were focused on the chemical reaction at the middle of the channel without any electrodes. These studies took advantage of the flat velocity profile at the center of the microchannel with large width to height aspect ratio, $\gamma=l_c/h$. They have been mainly focused on the binary first order reaction-diffusion equation which can be solved analytically \cite{Galfi1988}. Such models used in combination with chemical fields measurements obtained from fluorescence imaging techniques \cite{Baroud2003}, Raman spectroscopy \cite{Salmon2005} or infrared spectroscopy \cite{Chevalier2021,Chan2009} enabled accurate measurements of the reaction properties, i.e. the mass diffusivity and the reaction rate coefficient. These works have led to a better understanding of the reaction properties in laminar reactors. In contrast, few studies have been focused on the mass transfer at the electrode interface in the microfluidic chips, mainly due to non-constant velocity profile near the electrodes or the microchannel walls. \\

Close to the wall, the velocity profile in the microchannel is no longer flat due to the  no slip condition (the same assumption can be done close to the electrodes \cite{Tabeling2005}). In such cases, the mass transfer differs from the classical solution derived in the middle of the channel. Ismagilov et al. \cite{Ismagilov2000} showed that the expression of the diffusion boundary layer, $\delta_D$, along the channel length, $x$, follows a typical power law of one third, i.e. $\delta_D\propto x^{1/3}$, as theoretically predicted in the seminal works from Lévêque \cite{leveque1928lois}. The complete description of this phenomenon (mass diffusion with non-constant velocity profile) is known under the name of the Graetz problem \cite{Gervais2006}. Several solutions to this problem were found, using a parabolic laminar profile \cite{Ghez1978,Campo2020}. However, the case of mass transfer at the electrode's interface when the flow profile is not parabolic has been less studied. Bazant's group is one of the most active on this topic, publishing several works where it was shown how the current density distribution in microfluidic fuel cells was affected by the velocity profile \cite{Braff2013,Braff2013b,Biesheuvel2011}. Thus, the development of new mass transport analytical models focused on the exact velocity profile in the microchannel would achieve two goals. The first would be a strengthened understating of the reactant concentration distribution in energy conversion chips, and the second would be convenient measurements of the current density distribution at the channel/electrode interface from a known or measured concentration field.\\

A prime interest in the research of analytical or semi-analytical solutions of heat and mass transfer in microfluidic chips is to develop accurate measurement techniques of heat or mass flux density distributions. In the case of the mass transfer, the concentration distribution can be relatively well measured using several spectroscopic methods (FTIR, Raman, UV or visible spectroscopy). If an accurate model of the mass transfer is associated to these measurements, it is possible to estimate the flux density using an inverse processing method \cite{Brereton2000}. In particular, the formalism of transfer functions (in the terms of system analysis) is a very useful tool for performing the rapid and accurate inverse method \cite{Aouali2021}. Thus, with the development of a microfluidic electrochemical chip, research of mass transfer function between the concentration field and the mass flux at the electrode interface needs to be pursued toward the development of accurate inverse methods and new experimental characterization methods for those systems.\\

The present work aims to address this research of semi-analytical solutions of the mass transfer in a microfluidic electrochemical chip to derive the scaling and physical laws, and the system transfer function linking the mass flux and the concentration distributions. Considering a simplified but predictive averaged velocity profile in a large aspect ratio microfluidic channel, a semi-analytical solution in the Laplace domain is derived. It is also validated against a numerical model solving the complete problem and the Lévêque theory. The use of such a semi-analytical model would elucidate the mass transfer problem at the electrodes interface in regards to transfer functions of the current density and  concentration field in the microchannel. Such formalism would pave the way toward improved energy conversion chips by enabling important performance characterization tools, such as through accurate measurements of the microchannel current density distribution. The complete description of the model and the related assumptions are given in section 2, followed by the model validation and the transfer functions discussed in section 3.

\section{Mathematical modeling}

\subsection{General model}
\label{s_general_model}
\begin{figure}[hbtp]
\centering
\includegraphics[scale=.6]{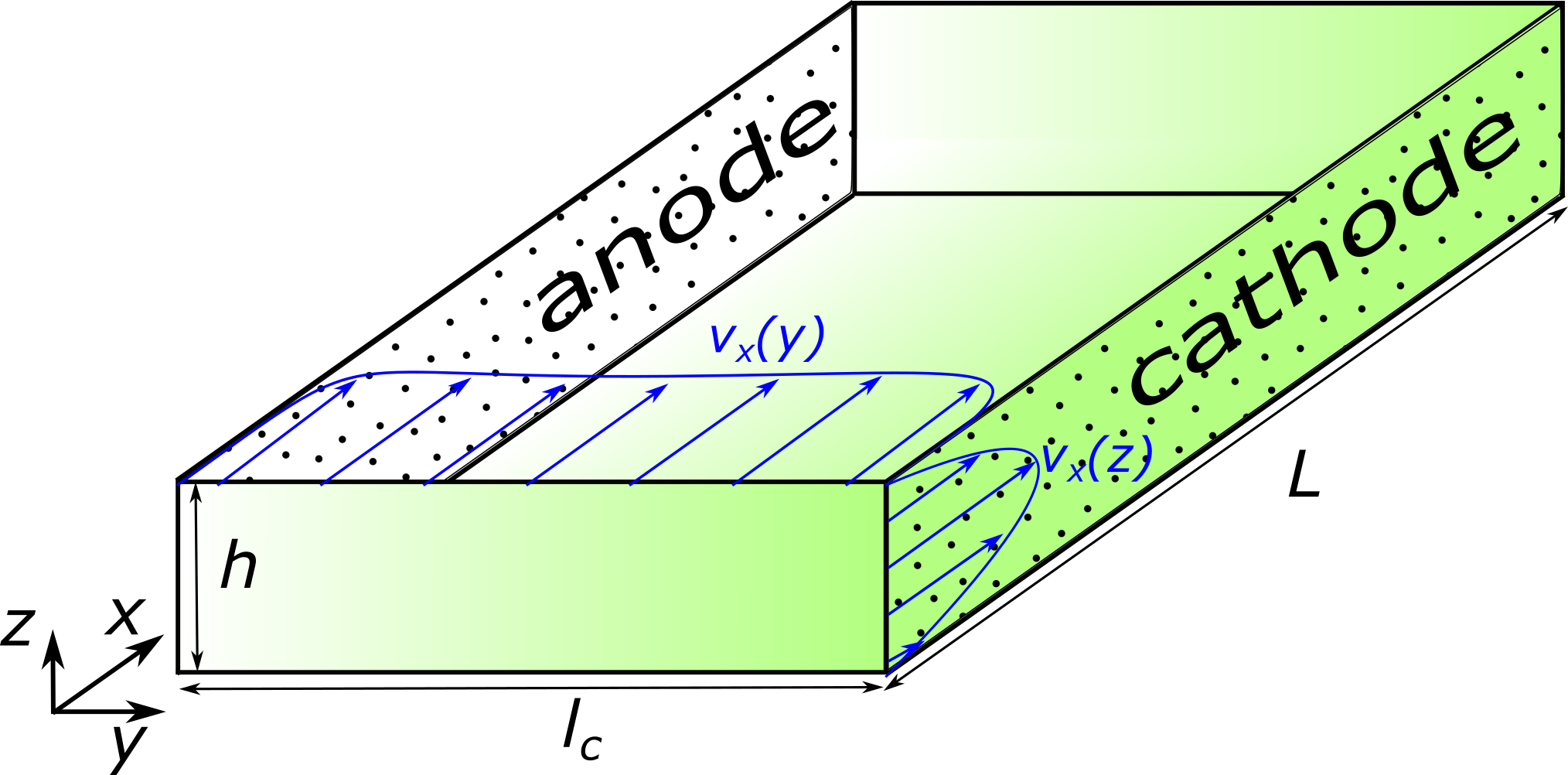}
\caption{Schematic of a microfluidic electrochemical device with electrodes on the side walls.}
\label{f_schema}
\end{figure}

The geometry of a basic laminar microfluidic electrochemical chip is schemed in Figure \ref{f_schema}. A large width to height aspect ratio channel is presented with electrodes on side walls. In such a configuration, the velocity profile is assumed to be laminar: quite flat in the y-direction and parabolic in the z-direction. We consider a first order chemical reaction of a species at one electrode, i.e. at the anode, which creates a concentration gradient mainly in the y- and x-direction (assuming that the electrode is homogeneous). The distance between the two electrodes is considered larger than the Debye length \cite{Dydek2011}, ensuring the electroneutrality of the solute. Finally, in the case of a diluted solution (which is frequent with aqueous solutions), the Fickian diffusion can be used to describe the transport of the species concentration in a microchannel as 

\begin{equation}
\vec{\nabla}\cdot(\vec{v}c)+\vec{\nabla}\cdot(-D\vec{\nabla}c)=0,
\label{e_diff_gene}
\end{equation}
where $\vec{v}$ is the velocity field, $c$ is the concentration field and $D$ the mass diffusivity. In the case of non porous, plane and homogeneous electrodes, the first order reaction at its surface is written as
\begin{equation}
-D\vec{\nabla}c\cdot\vec{y}=\pm \frac{j}{n_eF}=\pm kc,
\label{e_CL_gene}
\end{equation} 
where $j$ is the local current density distribution along the electrode, $n_e$ is the number of electrons transferred in the reaction, $F$ is the Faraday's constant, $c_0$ is the initial concentration of the solution, and $k$ is the reaction rate coefficient for the reaction at the electrode surface \cite{Braff2013}, assumed to be constant along the electrode surface.\\

Further assumptions were also made to simplify the previous equations. The chemical properties ($k$ and $D$) of the reaction are considered constant as long as the solution is diluted. The flow rate in the microchannel was considered large enough to ensure both a neglected diffusion in the x-direction and a small diffusion boundary layer close to the electrode. This condition ensures a semi-infinite channel in the y-direction and can be validated in terms of large Peclet number, i.e. $\text{Pe}=vl_c/D\gg 1$ where $l_c$ is the channel width. Even close to the electrodes where the velocity is small \cite{Ismagilov2000}, this condition should stand. However, the large Peclet number makes the channel flow working mostly in the Taylor-Aris regime where the parabolic velocity profile enhances the apparent diffusion coefficient in the flow direction. However, one may show that the diffusion in the x-direction (parallel to the flow) is still negligible compared to the transverse term, since the x-direction gradients are very weak. In this work, the transport of charge particles or colloids via diffusiophoresis \cite{ault_shin_stone_2018} or diffusioosmosis \cite{C8SM01360H} is not considered. The chemical reactants are assumed to be diluted homogeneous aqueous solutions.\\

Thus, given the above-mentioned assumptions, one can show that the concentration field can be described in 2 dimensions (2D) as the concentration gradients are mainly located in the y- and x-direction. In addition, the concentration fields should be averaged in the z-direction to take into account of the effect of the parabolic flow in the same direction. This ends up with an average 2D concentration field to be solved. It is also important to note that this 2D average concentration field would be the one experimentally measured from fluorescence or spectroscopic techniques \cite{Salmon2005,Chan2009} as they do not discriminate the signal in the z-direction. Therefore, in the case of the 2D averaged concentration field in a semi-infinite channel with a current density consumed at the electrode's interface, equations  \ref{e_diff_gene} and \ref{e_CL_gene} become:
 \begin{eqnarray}
\bar{v}_x(y)\frac{\partial \bar{c}}{\partial x}&=&D\frac{\partial^2 \bar{c} }{\partial y^2}, \label{e_eq_diff}\\
-D\left.\frac{\partial \bar{c}}{\partial y}\right|_{y=0}&=& -k\bar{c}(x,y=0), \label{e_CL_diff}\\
\bar{c}(x,y\longrightarrow \infty)&=& c_0,\\
\bar{c}(x=0,y) &=&c_0,
\end{eqnarray}
where $\bar{c}$ is the average 2D concentration field, and $\bar{v}_x(y)$ is the averaged velocity profile obtained from the Navier-Stokes equation. The boundary condition $\bar{c}(x,y\longrightarrow \infty)$ is a direct consequence of the semi-infinite assumption made in case of a large Peclet number. The initial condition $\bar{c}(x=0,y)$ indicates a uniform concentration at the channel inlet.\\

To clarify scalings and simplify the analysis, the governing equations for reactant transport are made dimensionless, by defining the Damköhler number as $\text{Da}=kl_c/D$, the dimensionless concentration as $\tilde{c}=\bar{c}/c_0$, and the dimensionless positions as $\tilde{y}=y/l_c$ and $\tilde{x}=x/x_e$, where $x_e=v^*l_c^2/D$ is the channel scaling length and $v^*$ is the effective average inlet velocity defined in next section. Therefore, the previous set of equations is rewritten as

\begin{eqnarray}
f(\tilde{y})\frac{\partial \tilde{c}}{\partial \tilde{x}}&=&\frac{\partial^2 \tilde{c} }{\partial \tilde{y}^2}, \label{e_eq_diff_adim}\\
\left.\frac{\partial \tilde{c}}{\partial \tilde{y}}\right|_{\tilde{y}=0}&=& \text{Da}\tilde{c}(\tilde{x},\tilde{y}=0), \label{e_CL_diff_adim}\\
\tilde{c}(\tilde{x},\tilde{y}\longrightarrow \infty)&=& 1,\\
\tilde{c}(\tilde{x}=0,\tilde{y}) &=&1,
\end{eqnarray}

where the function $f(\tilde{y})$ describes the evolution of the average velocity profile. It is obtained from the resolution of the Navier-Stoke equations.

\subsection{Velocity profile}
Under the assumption of laminar and fully developed flow, the Navier-Stokes equations in a microchannel are reduced to the following Poisson equation \cite{Pagitsas1986} :

\begin{equation}
\frac{1}{\gamma}\frac{\partial^2 v_x}{\partial \tilde{y}^2}+\frac{\partial^2 v_x}{\partial \tilde{z}^2} = \frac{\Delta ph^2}{\eta L},
\label{e_poisson}
\end{equation}
where $\Delta p/L$ is the linear pressure gradient, $\eta$ is the fluid viscosity, $\gamma=l_c/h$ is the width-height aspect ratio of the channel, and $\tilde{z}=z/h$ is the dimensionless channel height. Using the no slip conditions on each wall, this equation can be solved in terms of Fourier series as \cite{Mortensen2005,Bazant2016}

\begin{equation}
v_x(\tilde{y},\tilde{z}) = \frac{\Delta ph^2}{12\eta L}\sum_{n,odd}^\infty\frac{48}{(\pi n)^3}\left[1-\frac{\cosh((2\tilde{y}-1)\frac{n\pi}{2}\gamma)}{\cosh(\frac{n\pi}{2}\gamma)}\right]\sin\left(n\pi\tilde{z}\right).
\label{e_v_profile}
\end{equation}
The velocity profile averaged in the z-direction, defined as $\bar{v}_x(\tilde{y}) = \int_0^1v_x(\tilde{y},\tilde{z})d\tilde{z}$, leads to 

\begin{equation}
\bar{v}_x(\tilde{y}) = \frac{\Delta ph^2}{12\eta L}\left[1-\sum_{n,odd}^\infty\frac{96}{(n\pi)^4}\frac{\cosh((2\tilde{y}-1)\frac{n\pi}{2}\gamma)}{\cosh(\frac{n\pi}{2}\gamma)}\right].
\end{equation}
Since the pressure drop is usually unknown, the term $\frac{\Delta ph^2}{12\eta L}$ can be replaced by $v^*=v(1-2(\pi\gamma)^{-1})^{-1}\approx v(1-0.63/\gamma)^{-1}$ where $v$ is the average velocity in the microchannel computed simply from the flow rate as $v=q_v/(hl_c)$. The term $1-2(\pi\gamma)^{-1}$ in $v^*$ stems from the integration $\int_0^1\bar{v}_x d\tilde{y}$ in the case of a large aspect ratio, i.e. $\gamma>2$ (see the ref. \cite{Bruus2008} for the details of the calculation).\\
Thus, keeping the case where the aspect ratio is large enough (in practice $\gamma>2$), a good approximation of the velocity profile can be found using only the first order of equation \ref{e_v_profile}

\begin{equation}
\bar{v}_x(\tilde{y}) \approx v^*\left[1-\frac{96}{\pi^4}\frac{\cosh((2\tilde{y}-1)\frac{\pi}{2}\gamma)}{\cosh(\frac{\pi}{2}\gamma)}\right].
\end{equation}
Following the assumption of large Peclet number made in section \ref{s_general_model}, the diffusion boundary layer, $\delta_D$, is kept close to the electrode interface, allowing consideration of a semi-infinite channel. In these conditions, only the negative exponential of the hyperbolic cosine function can be used to describe the velocity profile. Finally, an approximation considering $96/\pi^4\approx 1$ leads to the following equation that describes the average flow profile in a semi-infinite microchannel
\begin{equation}
\bar{v}_x(\tilde{y}) \approx v^*\left(1-\exp\left(-\pi\gamma\tilde{y}\right)\right).
\label{e_v_profile_simp}
\end{equation}
In practice, this equation is valid as long as the diffusion boundary layer thickness is small, i.e. $\delta(\tilde{x})_D\leq l_c/4$, and in a microchannel with a large aspect ratio ($\gamma >2$). It can be also noted that equation \ref{e_v_profile_simp} verifies the correct average velocity, $v$, in the channel as: $2\int_0^{1/2}\exp(-\pi\gamma \tilde{y})d\tilde{y}= 2(\pi \gamma)^{-1}\approx 0.63/\gamma$.

\subsection{Analytical solution}
In order to solve the mass transfer between electrodes, equations \ref{e_eq_diff_adim} and \ref{e_v_profile_simp} are combined to together as
\begin{equation}
\left(1-\exp\left(-\pi\gamma\tilde{y}\right)\right)\frac{\partial \tilde{c}}{\partial \tilde{x}}=\frac{\partial^2 \tilde{c} }{\partial \tilde{y}^2}.
\label{e_eq_gene}
\end{equation}

The semi-analytical solution to equation \ref{e_eq_gene} is obtained using the Laplace transform performed on the $\tilde{x}$ variable. It is defined as
\begin{equation}
\hat{c}(\tilde{p},\tilde{y})=\int_0^{+\infty}(\tilde{c}(\tilde{x},\tilde{y})-1)e^{-\tilde{p}\tilde{x}}d\tilde{x},
\label{e_Laplace}
\end{equation}
where $\tilde{p}$ is the dimensionless complex Laplace parameter. It is linked to the complex Laplace parameter, $p$, as $\tilde{p}=pv^*l_c^2/D$. Using the transformation \ref{e_Laplace} on equations \ref{e_eq_gene} and \ref{e_CL_diff_adim} leads to the following set of equations:

\begin{eqnarray}
\frac{1}{\alpha^2}\frac{d^2 \hat{c}}{d \hat{y}^2}&=&(1-e^{-2\hat{y}})\hat{c}, \label{e_eq_diff_laplace}\\
\left.\frac{d \hat{c}}{d\hat{y}}\right|_{\hat{y}=0}&=& \frac{2\text{Da}}{\pi\gamma}\left(\hat{c}(p,\hat{y}=0)+\frac{1}{\tilde{p}}\right), \\
\hat{c}(\tilde{p},\hat{y}\longrightarrow \infty)&=& 0,
\end{eqnarray}

where $\alpha^2=4\tilde{p}(\pi\gamma)^{-2}$, and $\hat{y}=\pi\gamma\tilde{y}/2=\pi y/(2h)$. The previous system of equations has a solution in terms of Bessel function of the first kind and of order $\alpha$, $J_\alpha$, as

\begin{equation}
\hat{c}(\tilde{p},\hat{y})=\frac{J_\alpha(\alpha e^{-\hat{y}})}{\frac{\tilde{p}^{3/2}}{2\text{Da}}(J_{\alpha+1}(\alpha)-J_{\alpha-1}(\alpha))-\tilde{p}J_\alpha(\alpha)}.
\label{e_sol_ana_Da}
\end{equation}
In the case of very fast kinetics, $\text{Da}\longrightarrow \infty$, so the concentration at the electrodes surface drops to zero, i.e.  $\tilde{c}(\tilde{x},\tilde{y}=0)=0$, and solution \ref{e_sol_ana_Da} is reduced to

\begin{equation}
\hat{c}(\tilde{p},\hat{y})=-\frac{J_\alpha(\alpha e^{-\hat{y}})}{\tilde{p}J_\alpha(\alpha)}.
\label{e_sol_ana_c0}
\end{equation}
Finally, using an numerical inverse Laplace transform \cite{Toutain2011}, the average 2D concentration field $\tilde{c}(\tilde{x},\tilde{y})$ can be computed from equations \ref{e_sol_ana_Da} or \ref{e_sol_ana_c0}.

\section{Results and discussion}
\subsection{Validation of the assumptions}
Before the results of the present semi-analytical model are discussed, the assumptions made in the previous section are thoroughly validated against the full numerical computation of the concentration and velocity fields. First, the exact solution of the cross section of the velocity field (see equation \ref{e_v_profile}) is presented in Figure 2(a) with dimensionless coordinates and an aspect ratio $\gamma=3$. It shows a parabolic profile in the z-direction (in particular located at the middle of the channel). The gradients in the y-direction are, however, mostly located close to the wall (or electrode). In contrast these gradients  remain quite flat at the middle of the channel.\\

\begin{figure}[H]
\centering
\includegraphics[scale=.8]{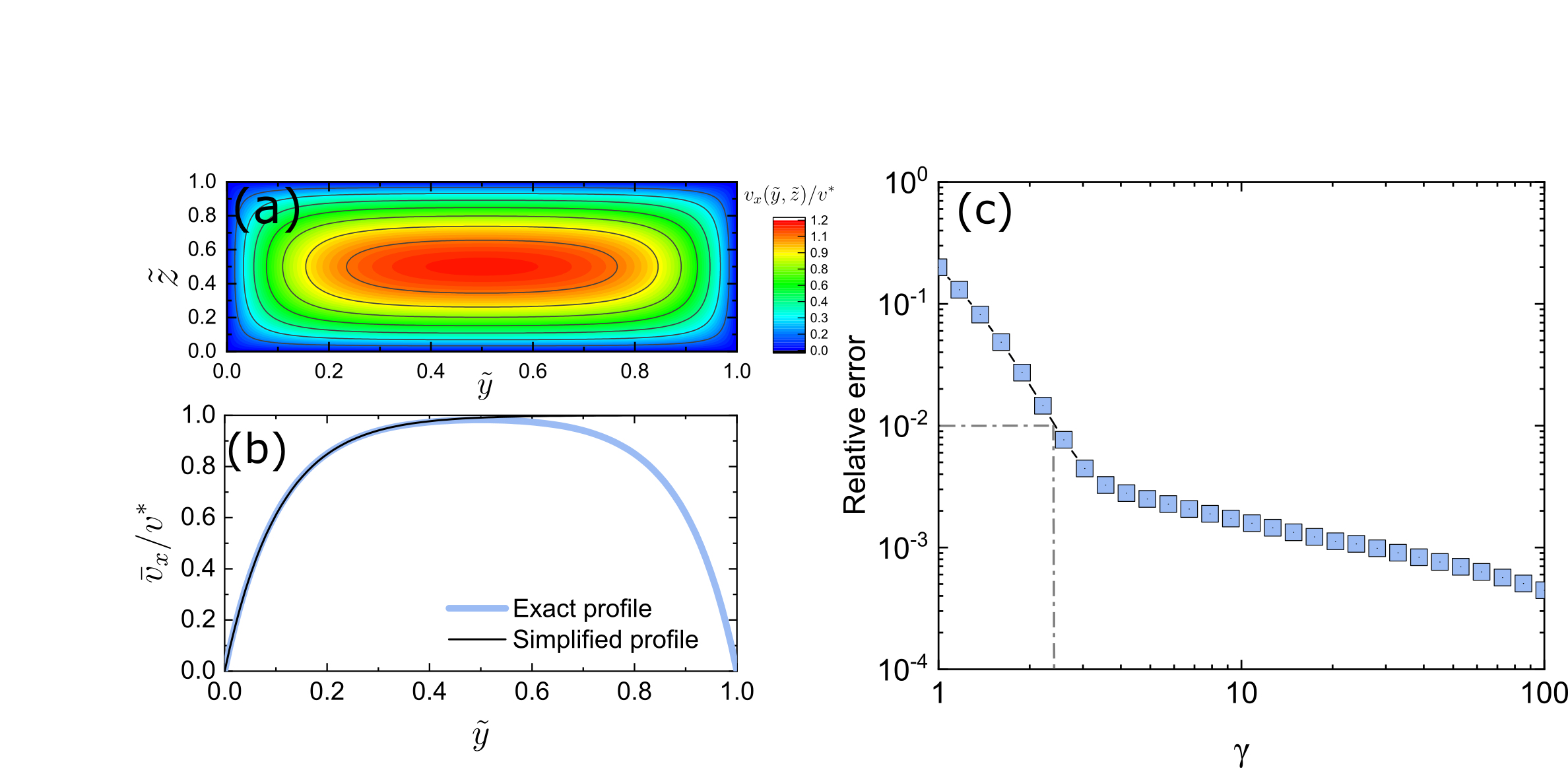}
\caption{Velocity field and profile for $\gamma=3$. (a), Solution of the 2D fully developed velocity field in a microchannel. (b), Averages of the exact and simplified velocity profiles. (c), Relative error between the exact and the simplified expression of the velocity profile for a range of aspect ratios, $\gamma$.}
\end{figure}

To illustrate the previous observation, the exact average velocity profile in the y-direction was computed and presented in Figure 2(b) for the same aspect ratio. The comparison with the simplified velocity profile (equation \ref{e_v_profile_simp}) is also shown, and an excellent agreement can be observed. As expected, equation \ref{e_v_profile_simp} is valid as long as the aspect ratio is larger than 2. This result is confirmed by computing the relative error between the averaged exact and simplified velocity profiles between $\tilde{y}=0$ and 0.5, for a range of aspect ratios, $\gamma$. This result is shown in Figure 2(c), and a relative error lower than 1\% is obtained when $\gamma>2.5$ (see the dashed lines). At $\gamma=2$, the relative error is about 3\% which can still be considered acceptable.\\

Following the same methodology, the present semi-analytical model was compared to the exact solution of the complete problem. This exact solution was computed from the average 2D concentration fields (equation \ref{e_eq_diff} with the associated boundary conditions and the exact velocity profile, equation \ref{e_v_profile}). It is solved numerically using a finite difference scheme and the Runge-Kutta algorithm. A no flux boundary condition on the concentration field was used at the opposite wall ($\tilde{y}=1$) in the numerical solution. The analytical solution was computed using a numerical inverse Laplace transform based on the Stehfest algorithm \cite{Toutain2011}. All the calculations were done using an aspect ratio of 3.\\

The results comparing the numerical and semi-analytical solution are summarized in Figure 3 for a range of y-position in the channel. In Figure 3(a), slow kinetics were chosen at the electrode interface, i.e. Da = 1. The dimensionless concentration profiles along the channel are observed to decrease along the channel position $\tilde{x}$ due to the chemical consumption at the electrode interface. In addition, the concentration is lower close to the electrode ($\tilde{y}=0$) and increases in the middle of the channel, i.e. $\tilde{y}=0.5$. This effect is due to the diffusion of the species \cite{Salmon2005}. Overall, an excellent agreement between the numerical and semi-analytical model is found as long as $\tilde{x}<0.2$. At higher $\tilde{x}$ values, the semi-infinite assumption is no longer valid, because the boundary diffusion layer becomes too large. In practice, for given microchannel dimensions and a specific species, one can compute the minimum average velocity to ensure that the diffusion in a semi-infinite channel follows
\begin{equation}
v>\frac{LD\left(1-0.63/\gamma\right)}{0.2l_c^2}.
\end{equation}

In Figure 3(b), similar results are presented in the case of fast kinetics at the electrode interface (modeled by a zero concentration at $\tilde{y}=0$). In such case, the concentration drops significantly faster in the microchannel as expected. An excellent agreement between the numerical model and the semi-analytical model (equation \ref{e_sol_ana_c0}) is also observed, and the same limit in terms of velocity can be drawn to ensure the semi-infinite assumption. Thus, all the results presented in Figure 3 validated and fixed the limit of the assumptions that were made to derive the semi-analytical solution of the mass transfer in large aspect ratio microfluidic electrochemical chips with electrodes on side walls.

\begin{figure}[H]
\centering
\includegraphics[scale=1]{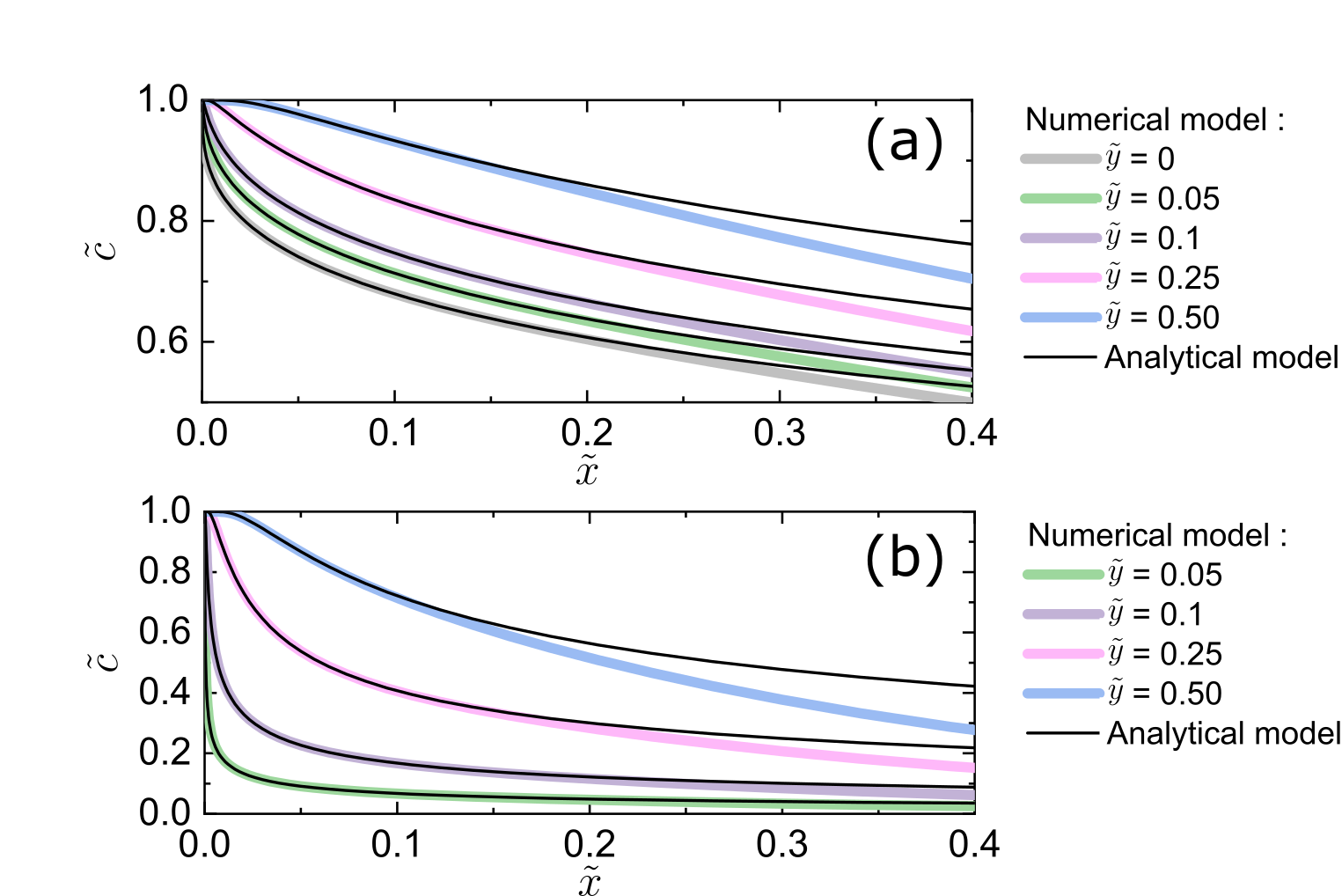}
\caption{Concentration profiles along the channel for a range of y-positions. (a), Slow kinetics, i.e. Da = 1. (b), Fast kinetics, i.e. $c(x,y=0)=0$. }
\end{figure}

\subsection{Comparaison with Lévêque's theory}

Moving forward with validation of the present model, and in particular concerning the assumption of negligible diffusion in the x-direction close to the electrode interface, a comparison with Lévêque's theory is presented. In his work, Lévêque \cite{leveque1928lois} obtained an approximation for the diffusion boundary layer in viscous shear flow at a no-slip surface. He obtained an exact solution widely used in theories of heat and mass transfer by forced convection and electrodialysis, and his model was widely validated experimentally with the case of heat transfer \cite{Campo2020}, but also with the case of mass transfer \cite{Ismagilov2000}.\\

Lévêque showed that the velocity profile can be linearised as $\bar{v}_x(\tilde{y})\approx \pi\gamma\tilde{y}v^*$ which is obtained by expanding Equation \ref{e_v_profile_simp} into a Taylor series and only retaining up to the linear term. This linearisation can be done when the Peclet number is quite large (i.e. Pe $> 1000$), which means that $\delta_D$ is very thin. This assumption leads to the following equation governing the mass transfer at the electrode interface
\begin{equation}
\pi\gamma\tilde{y}\frac{\partial \tilde{c}}{\partial \tilde{x}}=\frac{\partial^2 \tilde{c} }{\partial \tilde{y}^2}. \label{e_eq_diff_leveque}
\end{equation}
Using a constant concentration profile at the electrode interface ($c(\tilde{x},\tilde{y}=0)=0$), Equation \ref{e_eq_diff_leveque} can be solved in terms of an incomplete Gamma function \cite{Ghez1978} as
\begin{equation}
\tilde{c}_{Lev}(\tilde{x},\tilde{y})=\Gamma\left(\frac{\pi\gamma}{9}\frac{\tilde{y}^3}{\tilde{x}};\frac{1}{3}\right)
\end{equation}
where $\Gamma(a;b)$ is the incomplete Gamma function. This equation is valid in practice when the boundary diffusion layer is no larger than $\tilde{\delta}_D^{max}(\tilde{x})\sim(3\pi\gamma)^{-1}$.\\

In Figure 4, the comparison between Lévêque's theory and the present semi-analytical model for an aspect ratio of 3 is presented. The concentration profiles were computed in the y-direction for a range of channel positions. A good agreement between the model and Lévêque's theory is observed as long as the dimensionless diffusion boundary layer is kept lower than 0.04, which corresponds roughly to the maximum boundary layer $\tilde{\delta}_D^{max}$ indicated for $\gamma=3$. An important consequence of this validation, is that the assumption of negligible diffusion in the x-direction close to the electrode is valid as the Lévêque's theory and the diffuse behavior experimentally observed is reproduced closely.

\begin{figure}[H]
\centering
\includegraphics[scale=1]{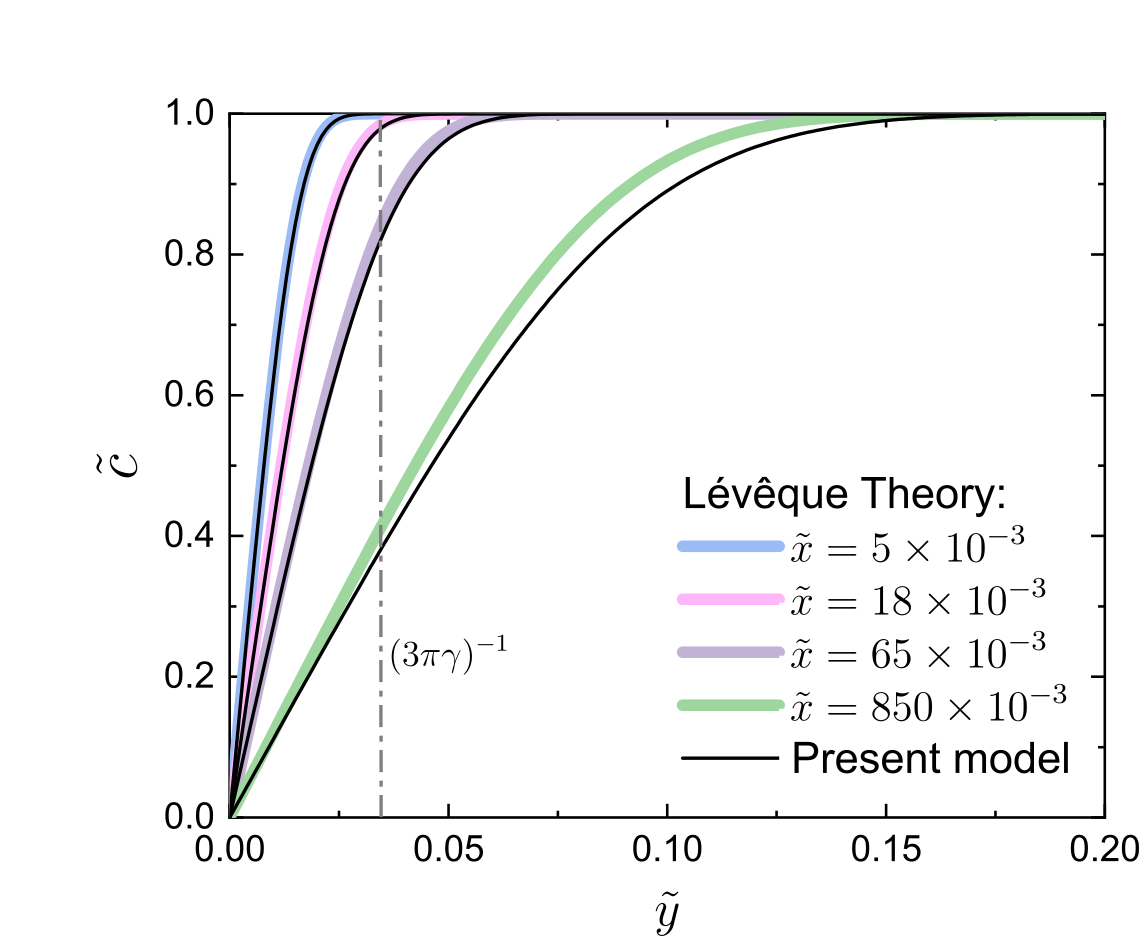}
\caption{Comparison and validation of the model with Lévêque's theory. The vertical dashed line represents the validity limit of Lévêque's theory in case of $\gamma=3$.}
\end{figure}

\subsection{Evolution of the diffusion boundary layer thickness}
An interesting parameter of mass transfer in the microchannel is the diffusion rate. In many applications this rate needs to be carefully controlled to enhance or reduced the mixing between products. A good indicator of the diffusion rate is the evolution of the 99\% diffusive boundary layer thickness along the channel (defined as the thickness where $\tilde{c}(\tilde{y})<0.99$). In absence of walls (or electrodes), it is known that this diffusive layer increases as the square root along the channel position, i.e. $\tilde{\delta}_D(\tilde{x})\propto \tilde{x}^{1/2}$ \cite{Tabeling2005}. This corresponds to the case where the exponential term from Equation \ref{e_v_profile_simp} vanishes at a large $y$ distance away from the wall. In contrast, the low velocity close to the walls (or electrodes) was found to lower the diffusion rate in the y-direction \cite{Ismagilov2000}, and a one third power law is usually expected, i.e. $\tilde{\delta}_D(\tilde{x})\propto \tilde{x}^{1/3}$. \\

In Figure 5, the 99\% diffusive boundary layer thickness is presented. It was computed from equation \ref{e_sol_ana_c0} for a microchannel with an aspect ratio of 7 to strengthen the effect, but the following conclusions can be applied to any aspect ratio. In the logarithmic scale, the slopes of the diffusion boundary layer close to the electrode and at the middle of the channel were estimated using linear regression on the first and last $\tilde{x}$ positions, respectively. As expected, the $\frac{1}{2}$ and $\frac{1}{3}$ slopes are obtained, confirming the good behavior of the proposed model. Furthermore, although the Lévêque's theory was only able to predict the $\frac{1}{3}$ slope close to the electrode, the model presented in this study bridges the gap between the diffusion from the electrode toward the middle of the channel. Therefore, it can be used to describe the mass transfer phenomenon at any position in the channel as long as the flow velocity is large enough to ensure semi-infinite diffusion.

\begin{figure}[H]
\centering
\includegraphics[scale=1]{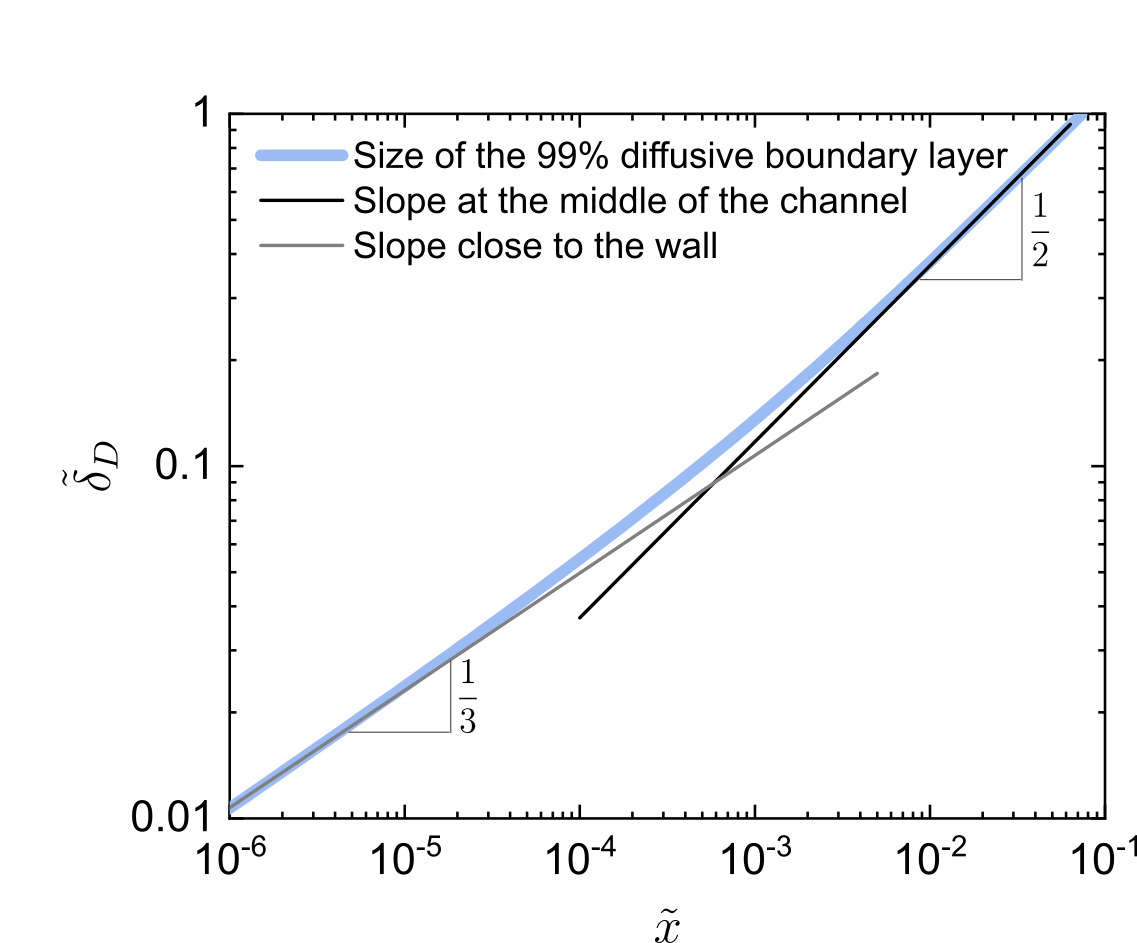}
\caption{Evolution of the 99\% diffusion boundary layer thickness along the channel position.}
\end{figure}

\subsection{Transfer function of the mass diffusion in a microchannel}
One direct application of these works is the use of equations \ref{e_sol_ana_Da} and \ref{e_sol_ana_c0} as transfer functions to measure the concentration or current density at the electrode interface. Such formalism is very well adapted to perform experimental measurements of concentration fields, or to compute the effect of a boundary condition on the concentration fields at a particular position in the microchannel without the use of any numerical partial differential equation solvers\\

Thus, the concentration distribution along the electrode using this formalism can be written in terms of convolution product as
\begin{equation}
\tilde{c}(\tilde{x},\tilde{y})=1+(\tilde{c}_0(\tilde{x})-1)\otimes\mathcal{H}_c(\tilde{x},\tilde{y}),
\label{e_conv_conc}
\end{equation}

where $\otimes$ indicates a convolution product, and $\mathcal{H}_c$ is the impulse transfer function for a Dirac concentration profile, i.e. $\tilde{c}(\tilde{x},\tilde{y}) = \delta(\tilde{x})$, where $\delta(x)$ is the Dirac distribution. In the Laplace domain, $\mathcal{H}_c$ is defined as
\begin{equation}
\hat{\mathcal{H}}_c(\tilde{p},\tilde{y})=\frac{J_\alpha(\alpha e^{-\pi\gamma\tilde{y}/2})}{J_\alpha(\alpha)}.
\end{equation}
Thus using an inverse numerical Laplace algorithm (such as Stehfest \cite{Toutain2011}), one can compute $\mathcal{H}_c=\mathcal{L}^{-1}\lbrace\hat{\mathcal{H}}_c\rbrace$, and use it to deconvolute the concentration field, which enables the measurement of the concentration profile at the electrode interface.\\

More interestingly, the same process can be done to compute the current density distribution along the electrode
\begin{equation}
\tilde{c}(\tilde{x},\tilde{y})=1+\tilde{j}(\tilde{x})\otimes\mathcal{H}_j(\tilde{x},\tilde{y}),
\label{e_conv_current}
\end{equation}
where $\tilde{j}(\tilde{x})=j(\tilde{x})l_c/(n_eFDc_0)$ is the dimensionless current density distribution along the electrode, and $\mathcal{H}_j$ is the impulse transfer function based on the distribution of the Dirac concentration flux, i.e. $\left. d\tilde{c}/d\tilde{y}\right|_{\tilde{y}=0}=\delta(x)$. In the Laplace domain, this transfer function is defined as:
\begin{equation}
\hat{\mathcal{H}}_j(\tilde{p},\tilde{y})=\frac{2J_\alpha(\alpha e^{-\pi\gamma\tilde{y}/2})}{\sqrt{\tilde{p}}\left(J_{\alpha+1}(\alpha)-J_{\alpha-1}(\alpha)\right)}.
\end{equation}
Thus, using equations \ref{e_conv_conc} or \ref{e_conv_current} and the associated transfer function, one can measure the electrode chemical concentration distribution, $\tilde{c}_0(\tilde{x}$), or the current density distribution, $\tilde{j}(\tilde{x}$), along the channel from $\tilde{c}(\tilde{x},\tilde{y})$ (valid if the aspect ratio is large, i.e. $\gamma>2$). Such an approach used in combination with spectroscopic measurements (infrared, Raman,...) to measure the concentration field opens a new method for characterization of the electrode current density distribution in microfluidic electrochemical chips.\\

The only drawback to the proposed application is the need to know the mass diffusivity of the chemical in the microchannel. This problem can be solved using analytical methods such as those developed by Salmon et al. \cite{Salmon2005} or Baroud et al. \cite{Baroud2003}. In their works, they imaged the diffusion of two chemicals in the middle of the channel where classical analytical solutions exists. Thus, by combining all these methods, the complete mass transfer problem in microfluidic electrochemical chips can be thoroughly characterized to quantitatively assess their performance and eventually design improvements.

\section{Conclusion}
The mass transfer at the electrode interface in a microfluidic channel was studied in the case of a large width to height aspect ratio. With this condition, it was possible to write a semi-analytical model in the Laplace domain to model the 2D average concentration distribution. All the assumptions made to derive such model were carefully validated using both the complete numerical solution of the problem, and the analytical theory developed by Lévêque. In particular, it was shown that the present semi-analytical model is valid as long as the aspect ratio is larger than 2 and in the case of a large Peclet number to ensure the semi-infinite assumption.\\

The main outcome of this work is the semi-analytical expressions of the 2D average concentration fields (equations \ref{e_sol_ana_Da} and \ref{e_sol_ana_c0}) which enable a thorough description of the mass transfer close to the electrode (power law of $\frac{1}{3}$) and in the middle of the channel (power law of $\frac{1}{2}$). The concentration fields computed in this study correspond to the measurements performed in a microchannel using imaging spectroscopic techniques (infrared or Raman) or fluorescence techniques. Thus, the combination of the present model, written in terms of a transfer function, would enable the measurement of the concentration distribution or the current density distribution along the electrode interface. These results are of prime importance for performance characterization that will lead to design improvements of the next generation of microfluidic electrochemical chips.

\section*{Acknowledgment}
The author gratefully acknowledges the French National Research Agency (ANR) for its support through the project I2MPAC, grant number ANR-20-CE05-0018-01. Mr K. Krause and Dr. J. Maire are gratefully acknowledged for their fruitful discussions during the manuscript preparation.

\clearpage







\end{document}